\documentstyle[preprint,prb,aps]{revtex}
\begin{document}
\draft
\title{Thermal phase diagram of a model Hamiltonian
       for columnar phases of liquid crystals}
\author{M. H\'{e}bert and M.L. Plumer}
\address{Centre de Recherche en Physique du Solide and
D\'{e}partement de Physique}
\address{Universit\'{e} de Sherbrooke, Sherbrooke,
Qu\'{e}bec, Canada J1K 2R1}

\date{August 1995}
\maketitle
\begin{abstract}
We present the phase diagram and critical properties of a coupled $XY$-Ising
model on a triangular lattice using the mean-field approximation, the
Migdal-Kadanoff scheme of renormalization group and Monte-Carlo simulations.
The topology of the phase diagram is similar for the three techniques, with
the appearance of a phase with $XY$ order and Ising disorder.  The results
suggest a line of transitions belonging to the 2D-Ising universality class
in contrast with previous data indicating a new universality class.  This
model is relevant to the columnar phases of discotic liquid crystals [such
as hexa(hexylthio)triphenylene (HHTT)] in the limit of weak intercolumn
coupling.
\end{abstract}

\pacs{PACS : 61.30.Cz, 64.70.Md, 64.60.Ak, 75.10.Hk}
\section{INTRODUCTION}
\label{sec:intro}

Columnar phases of discotic liquid crystals are becoming a field of
increasing experimental and theoretical activity in areas traditional
to magnetic systems (modulated phases, stability, critical
phenomena, etc.) [\onlinecite{pg93,caille95a}].  Particular attention has
been given to hexa(hexylthio)triphenylene (known as HHTT) compounds since
they were found to be excellent photoconductors [\onlinecite{adam94}].
The photoinduced charge carrier mobility is four orders of magnitude higher
than organic materials suitable for electronic devices and opens the way to
improved conventional and novel applications.  At present, our efforts are
focused on the understanding of the positional and orientational
orderings of the columnar phases.

The HHTT compound, first synthesized in 1984 [\onlinecite{kohne84}], has
been thoroughly studied by X-ray diffraction
[\onlinecite{fontes89,idziak92,fontes88,heiney89}].  These measurements
indicate that two intermediate columnar phases are present between the low
temperature crystalline $K$ phase and the high temperature isotropic liquid
phase $I$.  Upon cooling from the $I$ phase, the system enters the
disordered hexagonal $D_{hd}$ phase at $T \simeq 93^\circ C$ where
the disk-like molecules form columns, which are arranged on a
triangular lattice.  The $D_{hd}$ phase is characterized by intracolumnar
positional and orientational short-range order and long-range order in the
triangular positions of the columns.  As the temperature is reduced even
further, the HHTT compound enters the ordered hexagonal columnar phase ($H$
phase) at $T \simeq 70^\circ C$.  In this phase, the molecules within the
columns show long-range positional and orientational order (resolution
limited [\onlinecite{fontes88}]) while maintaining the triangular
arrangement of columns.  Inside the columns, the molecules of $D_3$ internal
symmetry are equally spaced and rotated by an angle $\alpha$ of
approximately $45^\circ$ along the column axes [Fig. \ref{fig1}(a)].  The
columns can be either left-handed or right-handed.  As noted by Fontes
[\onlinecite{fontes89}], these observations leave only three degrees of
freedom for each column: the helicity of the column, the overall
orientational phase angle of the column and the vertical displacement of the
column.  The proposed [\onlinecite{heiney89}] structure of the $H$ phase of
HHTT is characterized by the formation of a $\sqrt{3} \times \sqrt{3} R
30^\circ$ superlattice [Fig. \ref{fig1}(b)] in the helicity pattern and
vertical positions of the molecules described as follows.  One third of the
columns are displaced by half an intracolumnar intermolecular distance in
the $\hat{z}$ direction (along the stacking axis).  This is
thought to be a mechanism to relieve part of the frustration due to steric
hindrance.  The displaced columns also have opposite helicity to the
undisplaced ones.

Other experiments (Rayleigh scattering [\onlinecite{gharbia92}]) on similar
compounds have served to elucidate the question of the stability
[\onlinecite{pg93}] of the $H$ phase.  It is currently advocated that the
$H$ phase of HHTT is characterized by quasi long-range order within the
columns [\onlinecite{caille95b}].  Orientational order in the low
temperature region of the $H$ phase has been explored by ground state
calculations [\onlinecite{hebert95b}], and reproduces experimental
observations [\onlinecite{fontes89}].

A model has been proposed recently to reproduce the sequence of phases of
the HHTT compound [\onlinecite{hebert95a}].  This simple model Hamiltonian
incorporates, to a certain extent, what was noted by Fontes
[\onlinecite{fontes89}] and mentionned earlier.  It is a generalization of a
model introduced by Plumer {\it et al} [\onlinecite{plumer93}].  The columns
are taken as rigid helices inscribed in a two-dimensional triangular array.
The state of a column is determined by its helicity $K_i$ (Ising-like
variable) and a global phase angle $\theta_i$ ($XY$-like variable).  The
vertical displacement of a column is then viewed as a change in the global
phase angle $\theta_i$: the displacement of a column in the $\hat{z}$
direction by half an intracolumnar intermolecular spacing is replaced by the
addition of the angle $\alpha/2$ to $\theta_i$ [\onlinecite{hebert95a}].
This simplifying approximation neglects fluctuations in positional and
orientational degrees of freedom of individual molecules and mimicks the
change in helicity of a column, which would result from a gradual unwinding
of the helix, by a simple flip in the variable $K_i$.  This model is thought
to be valid in the strong intracolumn coupling limit of the exact
interaction energy [\onlinecite{hebert95b}].

To second order in the moments of the mass density of the columns, the
interaction energy of the system, coupling $XY$ and Ising variables on a
triangular lattice, is [\onlinecite{hebert95a}]
\begin{eqnarray}
{\cal H}&=&-\frac{J'}{2} \sum_{\langle ij \rangle} (1 + K_i K_j)
\cos(\phi_i - \phi_j) \nonumber\\
&&-\frac{G'}{2} \sum_{\langle ij \rangle } (1 - K_i K_j) \cos(\phi_i +
\phi_j)
\label{eq1}\end{eqnarray}
with $K_i = \pm 1$ and $\phi_i = 3 \theta_i$ ($0 \leq \phi_i \leq 2 \pi$).
(We use primed parameters to be consistent with earlier calculations
[\onlinecite{hebert95a}] and to leave unprimed parameters for the exact
interaction energy [\onlinecite{hebert95b}].)  Note that the sign of $G'$ is
not relevant since changing $G' \rightarrow -G'$ and $\phi_i \rightarrow
\phi_i + \pi/2$ leaves (\ref{eq1}) invariant.  This Hamiltonian with $G' =
0$ has been studied extensively in relation to the fully frustrated $XY$
models and Josephson junction arrays in a transverse magnetic field
[\onlinecite{lee91a,granato91}].  They reported a single continuous
transition from an $XY$ and Ising ordered phase to a totally disordered
phase with exponents $\nu = 0.83(4)$ and $\beta/\nu = 0.14(2)$ associated
with the Ising order parameter [\onlinecite{lee91b}].  These results
suggested the possibility of a new universality class.  We note also that a
similar model with no Ising variables [Hamiltonian (\ref{eq1}) with $K_i =
K_j = 0$] has been studied recently [\onlinecite{horigushi94}].

It is also of interest to summarize briefly the intrinsic properties of
Hamiltonian (\ref{eq1}) in the low temperature portion of the phase diagram
[\onlinecite{hebert95a}].  For $G' = 0$, the effective exchange parameter
for $K_i K_j$ brings true long-range ferromagnetic order in the Ising
variables, while the $XY$ variables should exhibit Kosterlitz-Thouless
order.  For $J' = 0$, the effective exchange parameter for $K_i K_j$ is
antiferromagnetic.  On a triangular lattice, this results in an Ising
disordered state at all temperatures [\onlinecite{wannier50}].  On the other
hand, the term $\cos(\phi_i+\phi_j)$ breaks the rotation invariance
permitting true long-range order in the $XY$ variables.  The competition
between the $J'$- and $G'$-terms are responsible for a rich $T - J'$ phase
diagram within mean-field theory [\onlinecite{hebert95a}].

In this work, we study the $T - G'$ phase diagram and critical behavior of
Hamiltonian (\ref{eq1}) for $J' = 1$.  In Section \ref{sec:mf}, the
mean-field phase diagram (following [\onlinecite{hebert95a}]) is presented
to give an indication of the type of order which may occur.  In Section
\ref{sec:rg}, we present the Midgal-Kadanoff renormalization group study of
Hamiltonian (\ref{eq1}).  Then, the phase diagram is calculated along with
the critical exponents from preliminary Monte-Carlo simulations (Section
\ref{sec:mc}).  Finally, general remarks and conclusions are drawn in
Section \ref{sec:conclu}.

\section{MEAN-FIELD PHASE DIAGRAM}
\label{sec:mf}

The mean-field phase diagram for $G' = -1$ has already been published
[\onlinecite{hebert95a}].  Here, we use the same technique to calculate the
$J' = 1$ phase diagram ($J' = 1$ fixes the energy scale).  The
main purpose of this calculation is to gain knowledge of the order
parameters describing each phase.  To sixth order in a Landau-type expansion
of the free energy for a unit cell (see Ref. [\onlinecite{hebert95a}]), the
resulting phase diagram is given Fig. \ref{fig2}.

Despite the fact that the mean-field approximation is crude, a lot of
information can be extracted from Fig. \ref{fig2}.  The low temperature
phase (phase III) is characterized by $XY$ and Ising order [$\langle
\phi_i\rangle = \langle \phi_j\rangle = \pi/2$ and $(\langle
K_1\rangle\:\langle K_2\rangle\:\langle K_3\rangle) = (+++)$].  Phase
II shows the existence of $XY$ order with Ising disorder
[$\langle\phi_i\rangle = \langle\phi_j\rangle \neq 0$ and $(\langle
K_1\rangle\:\langle K_2\rangle\:\langle K_3\rangle) = (000)$].  Finally, the
high temperature phase (phase I) has both $XY$ and Ising disorder.  As
expected, for $G' = 0$, we have a direct transition from phase III to phase
I [\onlinecite{lee91a}], but the order of the transition is wrong (see
below) within the mean-field approximation.  The $G'$-term is responsible
for the stabilization of phase II with $XY$ order and Ising disorder.  In
fact, this phase II appears only for $G' \agt 0.25$.  Note that italic
letters ($I$, $D_{hd}$, and $H$) refer to phases reported experimentally in
HHTT and roman numerals (I, II and III) denotes of the phases in the present
study.

\section{MIGDAL-KADANOFF RENOMALIZATION GROUP APPROACH}
\label{sec:rg}

In order to go beyond the Landau-type mean-field theory of Section
\ref{sec:mf} and include fluctuations of the order parameters, we apply the
Migdal-Kadanoff renormalization group scheme to a wider class of
Hamiltonians
\begin{eqnarray}
\beta{\cal H}_{RG}&=&-\frac{\beta J'}{2} \sum_{\langle ij \rangle} (1 + K_i
K_j) \cos(\phi_i - \phi_j) \nonumber\\
&&-\frac{\beta G'}{2} \sum_{\langle ij \rangle} (1 - K_i K_j) \cos(\phi_i +
\phi_j)\nonumber\\
&&-\frac{\beta L'}{2} \sum_{\langle ij \rangle} K_i K_j
\label{eq2}\end{eqnarray}
of which Hamiltonian (\ref{eq1}) is the $L' = 0$ limiting case.  The term $
-\frac{\beta L'}{2} \sum_{\langle ij \rangle} K_i K_j$ is added to
(\ref{eq1}) since its form is not preserved under renormalization
(the renormalization process creates $K_iK_j$ couplings).  The
Migdal-Kadanoff approach has been applied to Hamiltonian (\ref{eq2}) with
$G' = 0$ by a number of authors [\onlinecite{lee91a,li94}].  Those
calculations were carried out on a square lattice with rescaling factors $b
= 2$ and $b = 3$ in order to detect an antiferromagnetic Ising order at
large and negative $L'$.  The addition of the $G'$-term on a triangular
lattice, representing an intrinsic Ising frustration, is responsible for the
apparition of a new phase characterized by $XY$ order with Ising disorder,
and should be given special attention.  For these reasons, we will calculate
the phase diagram of (\ref{eq2}), for $G' < J'$, using the Migdal-Kadanoff
approximation implemented by Lee {\it et al} [\onlinecite{lee91a}], directly
on the triangular lattice with a rescaling factor $b = 2$.

To apply the Migdal-Kadanoff transformation, bonds are moved in such a way
that half the sites (to be integrated out) are linked to only two neighbors
(in one spatial direction).  This bond moving leads to a one-dimensional
decimation to obtain the effective exchange parameters between remaining
sites.  In terms of new variables $u(\theta) =\exp[U(\theta) - U(0)]$,
$v(\theta) =\exp[V(\theta) - V(0)]$ and $z = \exp[U(0) - V(0) + \beta L']$
with $U(\theta) = \beta J' \cos(\theta)$ and $V(\theta) = \beta G'
\cos(\theta)$, the recursion relations are
\begin{eqnarray}
\left[ u'(\phi_1 - \phi_2) \right]^2 & = & \frac{z^4 A_1(\phi_1 - \phi_2) +
z^{-4} A_2(\phi_1 - \phi_2)}{z^4 A_1(0) + z^{-4} A_2(0)}
\nonumber\\
\left[ v'(\phi_1 + \phi_2) \right]^2 & = & \frac{A_3(\phi_1 + \phi_2) +
A_4(\phi_1 + \phi_2)}{A_3(0) + A_4(0)}
\nonumber\\
\left[ z' \right]^2 & = & \frac{z^4 A_1(0) z^{-4} A_2(0)}{A_3(0) + A_4(0)}
\label{recur1}
\end{eqnarray}
where
\begin{eqnarray}
A_1(\theta) & = & \int_0^{2\pi} \frac{d \bar{\phi}}{2\pi}
u^4(\theta - \bar{\phi}) u^4(\theta) \nonumber\\
A_2(\theta) & = & \int_0^{2\pi} \frac{d \bar{\phi}}{2\pi}
v^4(\theta - \bar{\phi}) v^4(\theta) \nonumber\\
A_3(\theta) & = & \int_0^{2\pi} \frac{d \bar{\phi}}{2\pi}
u^4(\theta - \bar{\phi}) v^4(\theta) \nonumber\\
A_4(\theta) & = & \int_0^{2\pi} \frac{d \bar{\phi}}{2\pi}
v^4(\theta - \bar{\phi}) u^4(\theta).
\label{Ai}\end{eqnarray}
The primed variables in (\ref{recur1}) are the renormalized exchange
parameters (after decimation) and should not be confused with $J'$, $G'$ and
$L'$ which are the coupling constants of the original Hamiltonian
(\ref{eq2}).  In deriving (\ref{recur1}), the only assumption on the
potentials $U(\theta)$ and $V(\theta)$, during the renormalization process,
is their periodicities.

The phase diagram obtained by numerically iterating (\ref{recur1}) is
presented in Fig. \ref{fig3} for $J' = 1$ and $L' = 0$.  It is comprised of
four phases: I- a high temperature disordered phase, II- a phase with $XY$
order with Ising disorder, III- a low temperature fully ordered phase and
IV- a phase with Ising order with $XY$ disorder.  The low temperature phase
III is characterized by $XY$ and Ising order [$\beta J' \rightarrow (\beta
J')^\star$, $\beta G' \rightarrow (\beta G')^\star$ with $(\beta J')^\star =
(\beta G')^\star$, and $\beta L'\rightarrow \infty$].  The Hamiltonian with
$\beta G' = 0$ [\onlinecite{lee91a}] a is stable limit under the
renormalization group transformations (\ref{recur1}), as well as the $\beta
J' = 0$ limit.  Thus, in the $\beta G' = 0$ case, we get $\beta J'
\rightarrow (\beta J')^\star$, $\beta G' \rightarrow 0$ and $\beta
L'\rightarrow \infty$.  The intermediate temperature phase II shows $XY$
order with Ising disorder ($\beta J' \rightarrow (\beta J')^\star$, $\beta
G' \rightarrow (\beta G')^\star$ with $(\beta J')^\star = (\beta G')^\star$,
and $\beta L'\rightarrow 0$).   The PF line (Fig. \ref{fig3}) corresponds to
an Ising-like transition within this approximation.  In fact, putting
$(\beta J')^\star = (\beta G')^\star$ in (\ref{recur1}) and ({\ref{Ai}), we
find $A_i^\star (0) = A_j^\star (0)$ and
\begin{equation}
\left[z'\right]^2 = \frac{z^4 + z^{-4}}{2}\label{spin1/2}
\end{equation}
which is the Migdal-Kadanoff recursion relation for the spin-1/2 Ising model
with $b=2$.  The CPE line is relatively difficult to locate due to the high
temperature drift suffered by the Migdal-Kadanoff approximation.  The other
intermediate temperature phase IV is characterized by Ising order with $XY$
disorder ($\beta J' \rightarrow 0$, $\beta G' \rightarrow 0$ and $\beta
L'\rightarrow \infty$).  This phase is not seen in the mean-field
approach, and we argue that it is an artefact of the renormalization group
scheme.  In fact, by looking at Hamiltonian (\ref{eq1}) [or (\ref{eq2}] with
$L' = 0$), one is easily convinced that there cannot exist Ising order
without $XY$ order.  The PD line cannot be located precisely since the
Migdal-Kadanoff approach does not reproduce a line of true fixed points for
the $XY$ model.  Nevertheless, this scheme generates a line of almost fixed
points which gives a hint of $XY$-like ordered phases.  Finally, because of
these shortcomings of the Migdal-Kadanoff approximation, the exact topology
of the bifurcation point P (insets A and B of Fig. \ref{fig3}) cannot be
resolved.

As can be observed in Fig. \ref{fig3}, every phase transition is second
order.  The technique used to arrive at this conclusion is a generalization
of the procedure known to reproduce the first order transition of the
$q$-state Potts model for $q > 4$ [\onlinecite{nienhuis79}].  Our specific
implementation is due to Lee {\it et al} [\onlinecite{lee91a}].  The idea is
to have vacancies to prevent overestimation of the local order in the
system.

The lattice-gas version of (\ref{eq2}) reads
\begin{eqnarray}
\beta{\cal H}_{LG}&=&-\frac{\beta J'}{2} \sum_{\langle ij \rangle} (1 + K_i
K_j) t_i t_j \cos(\phi_i - \phi_j) \nonumber\\
&&-\frac{\beta G'}{2} \sum_{\langle ij \rangle} (1 - K_i K_j)
t_i t_j \cos(\phi_i + \phi_j)\nonumber\\
&&-\frac{\beta}{2} \sum_{\langle ij \rangle} \left( L' K_i K_j t_i t_j
+ K' t_i t_j\right) + \Delta \sum_i t_i,
\label{eq6}\end{eqnarray}
where the vacancy variable is $t_i = 0, 1$ and $\Delta$ is the fugacity.
Applying the same renormalization scheme to (\ref{eq6}), with $w = \exp[K' +
V(0) + U(0)]$ and $y = \exp(\Delta)$, we get the following recursion
relations
\begin{eqnarray}
\left[ u'(\phi_1 - \phi_2) \right]^2 & = & \frac{2+ \left[z^4 A_1(\phi_1 -
\phi_2) + z^{-4} A_2(\phi_1 - \phi_2)\right] w^4 y^{-2/3}}
{2+ \left[z^4 A_1(0) + z^{-4} A_2(0)\right] w^4 y^{-2/3}}
\nonumber\\
\left[ v'(\phi_1 + \phi_2) \right]^2 & = & \frac{2+ \left[A_3(\phi_1 +
\phi_2) + A_4(\phi_1 + \phi_2)\right] w^4 y^{-2/3}}
{2+ \left[A_3(0) + A_4(0)\right] w^4 y^{-2/3}}
\nonumber\\
\left[ w' \right]^2 & = & \frac{w^8 y^{8/3} \left\{ 2+ \left[z^4 A_1(0) +
z^{-4} A_2(0)\right] w^4 y^{-2/3} \right\}\left\{ 2+ \left[A_3(0) +
A_4(0)\right] w^4 y^{-2/3} \right\}}{2 w^{-2} y^{2/3} + B_1 z^2 + B_2 z^{-2}}
\nonumber\\
y' & = & \frac{w^{-12} y^6}{2 w^{-2} y^{2/3} + B_1 z^2 + B_2 z^{-2}}
\nonumber\\
\left[ z' \right]^2 & = & \frac{2+ \left[z^4 A_1(0) + z^{-4} A_2(0)\right]
w^4 y^{-2/3}}{2+ \left[A_3(0) + A_4(0)\right] w^4 y^{-2/3}}
\label{recur2}
\end{eqnarray}
with
\begin{eqnarray}
B_1 & = & \int_0^{2\pi} \frac{d \bar{\phi}}{2\pi}
u^4(\bar{\phi})\nonumber\\
B_2 & = & \int_0^{2\pi} \frac{d \bar{\phi}}{2\pi}
v^4(\bar{\phi}).
\label{Bi}\end{eqnarray}
The $\Delta t_i$-term has been distributed equally between bonds at site $i$
before moving them.  This is necessary to preserve the site density at each
step of the renormalization process.  The line of discontinuity fixed
points, signaling the first order transition, appears for $w, y, z
\rightarrow \infty$ [\onlinecite{lee91a,nienhuis75}].  In this limiting
case, an almost fixed line appears when
\begin{equation}
y' \left[ w' \right]^{-2} \left[ z' \right]^2 = \frac{y^{14/3}
w^{-12} z^{-12}}{A_1^\star(0)^2}.
\label{fixedline}
\end{equation}
The criterion to have a discontinuity fixed point (corresponding to a first
order transition) states that eigenvalue $\lambda$, for small deviations
from the fixed point along the unstable direction, must be given by $\lambda
= b^d$ ($d$ is the dimensionxnality of the lattice)
[\onlinecite{nienhuis75}].  We can see, from (\ref{fixedline}), that this is
certainly not the case.  Therefore, within the Midgal-Kadanoff
approximation and using the procedure of Nienhuis {\it et al}
[\onlinecite{nienhuis75}], the transitions in Fig. \ref{fig3} are continuous.

It is of interest compare our results with Lee {\it et al}
[\onlinecite{lee91a}] in the limit of $G' = 0$ and $L' = 0$.  Within the
Migdal-Kadanoff approximation, both calculations give the sequence of phases
III $\leftrightarrow$ IV $\leftrightarrow$ I with second-order transitions
between them.

\section{MONTE-CARLO SIMULATIONS}
\label{sec:mc}

In this section we report the results of a preliminary Monte-Carlo study of
the phase diagram and
critical behaviour of the model (\ref{eq1}).  We used the traditional
Monte-Carlo technique to locate the phase boundaries and to verify the
overall temperature behaviour of the thermodynamic quantities, especially
the order parameters and susceptibilities.  Then, we used the Monte-Carlo
histogram technique to locate accurately the transition temperatures and to
estimate critical exponents by finite-size scaling analysis.  Both
Monte-Carlo approaches used a sequential version of the Metropolis algorithm
to test separately the Ising and $XY$ trial configurations.  A Monte-Carlo
step (MCS) is as follows.  For each lattice site, the Metropolis test is
applied to the Ising variable ($K_i$) and then to the $XY$ variable
($\phi_i$).

\subsection{Monte-Carlo phase diagram}
\label{subsec:mcdia}

Fig. \ref{fig4} shows the resulting Monte-Carlo phase diagram.  It has
been obtained for a $L \times L$ lattice ($L = 36$) with $10^5$ MCS for
thermalization and $4 \times 10^5$ MCS for averaging.  We use the same
convention, as in the mean-field (Section \ref{sec:mf}) and renormalization
group (Section \ref{sec:rg}) phase diagrams, to name the phases (roman
numerals).  To describe the order present in each phase, we define the
following order parameters
\begin{eqnarray}
P_K & = & \left| \frac{1}{L^2} \sum_i K_i \right|\\
P_{Mj} & = & \left| \frac{3}{L^2} \sum_{i \in {\cal S}_j} \vec{S}_i \right|,
\label{op1}\end{eqnarray}
where ${\cal S}_j$ denotes one of the three sublattices and $\vec{S}_i =
(\cos(\phi_i) , \sin(\phi_i))$ are pseudo-spin variables.  The three
$XY$-like $P_{Mj}$ order parameters are very sensitive to sublattice
switching.  We therefore performed the simulations with the following
$XY$ order parameter,
\begin{equation}
P_M = \max_j (P_{Mj})
\label{op2}\end{equation}
which has smoother temperature dependence than $P_{Mj}$.  In order to
understand each phase in Fig. \ref{fig4}, we present the order parameters
$P_K$ and $P_M$ as functions of temperature [Fig. \ref{fig5}(a)].  Phase
III ($P_M$ and $P_K \neq 0$) is the phase with $XY$ and Ising order.  The
intermediate phase II keeps the $XY$ order ($P_M \neq 0$) but looses the
Ising order ($P_K = 0$).  At high temperatures, the isotropic I phase has
$XY$ and Ising disorder ($P_M$ and $P_K = 0$).  A similar behaviour to that
of the $XY$ order parameter $P_M$ and its susceptibility $\chi_M$ [Fig.
\ref{fig5}(b)] in the phase II has been reported for the six-state clock
model [\onlinecite{fujiki95}] and is characteristic of systems with
intrinsic disorder.

The general features of the Monte-Carlo phase diagram can be understood
easily.  For $G' = 0$, there is a single transition from phase III directly
to phase I [\onlinecite{granato91}].  This can be seen by noting that the
mechanism required to have a transition in the Ising variables is the
appearance of domain walls.  For $G' = 0$, at such a domain wall, the $XY$
variables are decoupled since $(1 + K_i K_j) = 0$.  In this case, Ising
disorder induces $XY$ disorder.  On the other hand, for $G' \gg J'$, there
should be no Ising order in the system (Section \ref{sec:intro}).  This is
seen in Fig. \ref{fig4}: the PQ$_1$ line of transitions goes toward
$T = 0$ as $G'$ gets larger.  Ultimately, for $G' \rightarrow \infty$,
there is only one transition from a phase with $XY$ order and Ising disorder
to the phase I.

\subsection{Finite-size scaling}
\label{subsec:scaling}

The rest of this Section is concerned with the finite-size scaling at the
four points (M, N, Q$_1$ and Q$_2$) shown in Fig. \ref{fig4}.  The goal
here is to extract only rough estimates of the critical exponents.  We are
also interested in detecting the order of the transitions and
the possible presence of a Kosterlitz-Thouless transition.  Finite-size
scaling using the Monte-Carlo histogram method has been recently reviewed by
Plumer {\it et al} [\onlinecite{plumer94}].  In our case, it involves the
specific heat ($C_v$), the energy cumulant ($U_e$), the order parameters
($P_M$ and $P_K$), their susceptibilities ($\chi_M$ and $\chi_K$), their
logarithmic derivatives ($V_M$ and $V_K$) [\onlinecite{peczak91}], and their
fourth-order cumulants ($U_M$ and $U_K$) [\onlinecite{binder81}].  We
performed a simulation for each lattice size $L = 54, 63, 72, 81, 90$
and $108$ at our best estimate of the transition temperature $T_o$ from the
results for the order parameters, such as in Fig. \ref{fig5}.  Typically, $2
- 3 \times 10^5$ MCS were used for thermalization and $6 - 10 \times 10^5$
MCS were kept for averaging.  Since we did only one simulation per lattice
sizes at $T_o$, the errors bars on the exponents are likely quite large
(typically 10 - 15 \%).

Let us focus on the transition M, at $G' = 0$ where $T_o = 2.175$ in
Fig. \ref{fig4}, to explain the details of the scaling analysis used here
and because these results will be compared with earlier published data
[\onlinecite{granato91}].  The well-studied fully frustrated $XY$ model is
also closely connected with Hamiltonian (\ref{eq1}) with $G' = 0$
[\onlinecite{yosefin85}].

To determine the critical temperature $T_c$, we used the scaling with
$L^{-1}$ (which assumes $\nu \simeq 1$) of the positions of the extrema of
the susceptibilities ($\chi_M$ and $\chi_K$) and of the logarithmic
derivatives ($V_M$ and $V_K$).  In the special case of $G' = 0$, the
cumulant-crossing method (using $U_K$) was also applied to get a better
estimate of $T_c$ [\onlinecite{peczak91}].  The results are summarized in
Fig. \ref{fig6}.  Both techniques give roughly the same $T_c$.  The scaling
of the extrema of $\chi_M$ and $V_M$
gives $T_{c-M} = 2.1631(10)$, while the scaling for
$\chi_K$ and $V_K$ gives $T_{c-K} = 2.1635(7)$.
Using the cumulant-crossing we get $T_{c-CC} = 2.1644(10)$.  From our data we
can conclude that, given the errors on the critical temperatures, the
transitions involving the $XY$-type and Ising-type order parameters occur at
the same temperature, as expected [\onlinecite{yosefin85,granato91}].  The
precision is quite good, especially when it is compared to the estimation of
$T_c$ in other systems [\onlinecite{plumer94b,reimers92}].  The value of
$T_c$ is required to find $\beta / \nu$ because $P_K \sim L^{-\beta/\nu}$ at
$T_c$.  The other exponents can also be found by a scaling at $T_c$, but in
our case the scaling of the extrema of the thermodynamical quantities proved
to be more precise.

The size dependence of the minimum of the energy cumulant ($U_e$) is given
in Fig. \ref{fig7} for the four transitions.  The fact that $U_e
\rightarrow 2/3$ when $L \rightarrow \infty$ indicates that the four
transitions are continuous [\onlinecite{challa86}].  This is in aggreement
with our renormalization group results (Section \ref{sec:rg}) and with the
results of Granato {\it et al} [\onlinecite{granato91}] in the case of $G' =
0$.

Fig. \ref{fig8} and \ref{fig9} show the scaling that leads to the
critical exponents $\nu_K$ and $\beta_K / \nu_K$ associated with the Ising
order parameter $P_K$.  In the case of $\beta_K / \nu_K$, we used the
critical temperature given by the scaling of the extrema of $\chi_K$,
$V_K$, $\chi_M$ and $V_M$.  The rough estimates are $\nu_K = 1.1(1)$ and
$\beta_K / \nu_K = 0.18(1)$.  The errors on the exponents are estimated
only by the goodness of the fit on the plots and do not account for the
(unknown) statistical error for each run.  From the relatively large scatter
of the points on those figures, it appears that there are large
statistical fluctuations.  Our estimates of the exponents are in apparent
disagreement with those of Granato {\it et al} [\onlinecite{granato91}].
Their study involved very long Monte-Carlo runs but on relatively small
lattice sizes ($L \leq 40$). Scaling with only our three smallest lattice
sizes ($L = 54$, $63$ and $72$), gives $\nu_K \simeq 0.74$ and $\beta_K /
\nu_K \simeq 0.10$, which are closer to their results.  It is clear that the
are significant finite-size effects.  Fig. \ref{fig10} presents the
scaling of $\chi_{Kmax}$ with $L$, giving $\gamma_K / \nu_K = 1.55(16)$.

These exponents ($\nu_K$, $\beta_K / \nu_K$ and $\gamma_K / \nu_K$) are
close to the 2D-Ising universality class ($\nu_K = 1$, $\beta_K / \nu_K =
0.125$ and $\gamma_K / \nu_K = 1.75$).  In this case, the specific heat
diverges like the logarithm of $L$.  Fig. \ref{fig11} shows $C_{vmax}$
against $\ln L$.  The straight line fit for $L \geq 63$ is a strong
indication that, contrary to what was suspected earlier
[\onlinecite{granato91}], the transition at $G' = 0$ is of the 2D-Ising
universality class.  Such logarithmic divergence of $C_{Vmax}$ for the
infinite lattice is also reported for the fully frustrated $XY$ model
[\onlinecite{lee86}].  Extensive Monte-Carlo simulations involving large
lattice sizes and better statistics are required to verify this conclusion.

Table \ref{tab1} lists the exponents for the four transitions considered,
along with $T_o$ and $T_c$.  The missing values for the exponents at points
Q$_1$ and Q$_2$ reflects the fact that for  $G' = 0.4$, the transitions
involve only Ising and $XY$ order respectively.  The errors on the exponents
are estimated only by the goodness of the fit.  The exponent $\alpha =
0\:(\log)$ means that $C_{vmax} \sim \ln(L)$, as in the 2D Ising model.  We
note that $\gamma_M / \nu_M$ at $G' = 0.4$ was estimated at $T_c$, because
$\chi_M$ shows only a broad maximum across the phase II and not a sharp peak
[see Fig. \ref{fig5}(b)].  It was impossible to get an estimate of
$\beta_K / \nu_K$ for transition Q$_1$, the data being too erratic.

Under the assumption that for $G' = 0.0$ (and $G' = 0.1$) there is only
a single transition, there can be only one caracteristic length that
diverges.  The exponents $\nu_K$ and $\nu_M$ are related to the divergence
at $T_c$ of the Ising type and $XY$ type correlation lengths, respectively.
Thus $\nu_K$ has to be equal to $\nu_M$, if there is a direct transition
from phase III to phase I.  This seems to be the case given the accuracy of
our results.  As noted in Ref. [\onlinecite{lee91a}], however, the Ising and
$XY$ transitions may be decoupled if $\nu = 1$.  This question has also been
addressed in relation to the antiferromagnetic $XY$ model on a stacked
triangular lattice [\onlinecite{plumer94b}].

Except for the values of $\beta_K / \nu_K$ at points N and Q$_1$, from the
data given in Table \ref{tab1} and Fig. \ref{fig7}, we tentatively
conclude that the transitions along the MPQ$_1$ line are continuous and
belong to the 2d-Ising universality class.  We exclude the possibility
of a transition of the Kosterlitz-Thouless type since at points M, N
and Q$_1$ (Fig.\ref{fig4}), $\nu_K \simeq \nu_M \simeq 1$ and $C_{vmax}$ is
proportional to $\ln(L)$ at $T_c$ [\onlinecite{bramwell94,bowen92}].  This
in agreement with our renormalization group treatment of model (\ref{eq1})
along the PF line of Fig. \ref{fig3} in Section \ref{sec:rg}.  As for the
other exponents, $\gamma_K / \nu_K$ and $\gamma_M / \nu_M$ seem to be stable
along the MPQ$_1$ line and suggest the 2D-Ising universality class.
However, the data for $\beta_K / \nu_K$ and $\beta_M / \nu_M$ are very noisy
(see Fig. \ref{fig9}) and no definite conclusions can be drawn.

The last transition studied (Q$_2$) is caracterized by the loss of $XY$
order while the Ising variables are disordered.  The specific heat scaling
shows no dependence on lattice size indicating that $\alpha \simeq 0$.
This rules out the possibility of a Kosterlitz-Thouless transition
[\onlinecite{teitel83}].  A similar behaviour has been observed in
frustrated antiferromagnetic $XY$ models in a magnetic field
[\onlinecite{lee86}].  In this case the exponents were reported to
be nonuniversal.  This might be the case here, but several points along the
PQ$_2$ would have to be done (with much better statistics) in order to
follow the evolution of the critical exponents.

\section{CONCLUSION}
\label{sec:conclu}

We have presented the phase diagram of a model for columnar phases of
HHTT in the limit of weakly interacting columns.  Its Hamiltonian couples
$XY$ and Ising variables.  We used the mean-field approximation, the
Migdal-Kadanoff scheme of renormalization group and Monte-Carlo
simulations.  These techniques give qualitatively the same topology
for the phase diagram, showing a phase with $XY$ order and Ising
disorder.  Renormalization group results and Monte-Carlo simulations
suggest that all that transitions are continuous.  From the renormalization
group analysis, we found that the transition from phase III to phase II
should be of the Ising type (Fig. \ref{fig3}).  Preliminary Monte-Carlo
simulations on large lattice sizes confirm this result and that the
transition from phase III directly to phase I is also of the 2D-Ising
universality class.  (We emphasize, however, that our statistical errors
appear to be quite large.)  This latter conclusion is in disagreement with
the suggestion by Granato {\it et al} [\onlinecite{granato91}] of a new
universality class.

\acknowledgements
We thank A. Caill\'{e} and A.M. Tremblay for useful discussions.
One of the authors (M. H.) wishes to thank H.T. Diep for discussions
on Monte-Carlo simulations and acknowledge the hospitality of the
Universit\'{e} de Cergy-Pontoise during the early stages of this work.  This
work was supported by the National Sciences and Engineering Research Council
(NSERC) of Canada, the Fonds pour la Formation des Chercheurs et de l'Aide
\`{a} la Recherche (FCAR) du Qu\'{e}bec and the Centre de Recherche en
Physique du Solide (CRPS).

%

%
\begin{figure}
\caption{(a)  Schematic side-view of a half period of HHTT in the
$H$ phase.  The full period is composed of 8 molecules of $D_3$ internal
symmetry symbolized by triangles.  The two helicities $K_i = \pm 1$ are
represented.
(b)  Structure of the $H$ phase in the basal plane showing the unit cell
(dashed line). Displaced columns are represented by open dots and
undisplaced columns by full dots.}
\label{fig1}
\end{figure}

\begin{figure}
\caption{Mean-field phase diagram of Hamiltonian (1) for $J' = 1$.
Different phases are noted by roman numerals.  Phase III is caracterized by
$XY$ and Ising order, phase II by $XY$ order with Ising disorder and phase I
by $XY$ and Ising disorder.  Helicity and angular configurations are
represented, and explained in the text.  Solid and dashed lines represent
first- and second-order phase transitions, respectively.  The square denotes
the location of critical end point.}
\label{fig2}
\end{figure}

\begin{figure}
\caption{Topological features of the phase diagram of Hamiltonian (2)
within the Mig\-dal-Kadanoff renormalization scheme for $J' = 1$ and $L' =
0$.  Phases are labelled as in Fig. 2, with phase IV representing $XY$
disorder with Ising order.  All transitions are continuous (dashed lines).
The structure at point P cannot be determined accurately; the two
possibilities are represented by the insets A and B.}
\label{fig3}
\end{figure}

\begin{figure}
\caption{Monte-Carlo phase diagram, with $J' = 1$, for a system size of
$36^2$ sites.  Different phases are reprensented as in Fig. 2.  Full circles
denote the transition temperatures for a given $G'$.  All transitions are
continuous.  Full squares and capital letters refer to the transitions where
finite-size scaling was performed.}
\label{fig4}
\end{figure}

\begin{figure}
\caption{(a) Temperature dependence of the Ising ($P_K$) and $XY$ ($P_M$)
order parameters for $G' = 0.4$.
(b) Temperature dependence of the susceptibilities associated with $P_K$ and
$P_M$.}
\label{fig5}
\end{figure}

\begin{figure}
\caption{Estimation of $T_c$ for $G' = 0$.  The solid lines shows the
scaling of the extrema of $\chi_K$, $V_K$, $\chi_M$ and $V_M$ vs $L^{-1}$.
The dashed lines represents the scaling of $T_c$ with respect to
$\ln^{-1}(L'/L)$ using the cumulant-crossing method.}
\label{fig6}
\end{figure}

\begin{figure}
\caption{Scaling of the minimum of the energy cumulant for the four
transitions considered.  Solid lines are guides to the eye.}
\label{fig7}
\end{figure}

\begin{figure}
\caption{Scaling of the logarithmic derivative of $P_K$.  The solid line
through the five largest lattice sizes gives the exponent $\nu_K = 1.1(1)$.
The dotted line is for the scaling using the two smallest lattice sizes
(see text).}
\label{fig8}
\end{figure}

\begin{figure}
\caption{Scaling of the order parameter $P_K$ at $T_c = 2.1635(7)$.  The
solid line through the five largest lattice sizes gives the exponent
$\beta_K / \nu_K = 0.18(1)$.  The dotted line is for the scaling using the
three smallest lattice sizes (see text).}
\label{fig9}
\end{figure}

\begin{figure}
\caption{Scaling of the susceptibility associated to $P_K$.  The solid line
through the five largest lattice sizes gives the exponent $\gamma_K / \nu_K
= 1.55(16)$.}
\label{fig10}
\end{figure}

\begin{figure}
\caption{Specific heat as a function of lattice size $\ln(L)$.}
\label{fig11}
\end{figure}

\narrowtext
\begin{table}
\caption{Summary of the critical exponents for the four transitions
considered.  The errors are estimated by the robustness of the fit
of the data and can be larger due to statistical fluctuations.  The
transitions Q$_1$ and Q$_2$ involve $XY$ and Ising orders, respectively, and
only those exponents are given.  Question marks indicate large scatter in
the data.}
\begin{tabular}{lcccc}
 &$G' = 0.0$ (M)&$G' = 0.1$ (N)&$G' = 0.4$ (Q$_1$)&$G' = 0.4$ (Q$_2$)\\
\tableline
$T_o$&2.175&2.165&1.745&2.37\\
$T_c$&2.1635(7)&2.148(2)&1.737(3)&2.376(5)\\
$\nu_K$&1.1(1)&1.18(16)&1.06(12)&-\\
$\beta_K / \nu_K$&0.18(1)&0.57(1)?& &-\\
$\gamma_K / \nu_K$&1.55(16)&1.58(16)&1.64(16)&-\\
$\nu_M$&0.90(8)&1.19(13)&-&1.09(7)\\
$\beta_M / \nu_M$&0.44(5)?&0.56(10)?&-&0.58(14)?\\
$\gamma_M / \nu_M$&1.46(6)&1.46(18)&-&1.10(14)\\
$\alpha$&0 (log)&0 (log)&0 (log)&0\\
\end{tabular}
\label{tab1}
\end{table}


\end{document}